\documentclass[aps,prb,twocolumn,superscriptaddress,floatfix]{revtex4}%
\usepackage{graphicx}
\usepackage{amsmath}
\usepackage{amsfonts}
\usepackage{amssymb}%
\def\nbZ {{\mathchoice {\hbox{$\sf\textstyle Z\kern-0.4em Z$}}
{\hbox{$\sf\textstyle Z\kern-0.4em Z$}} {\hbox{$\sf\scriptstyle
Z\kern-0.3em Z$}}  {\hbox{$\sf\scriptscriptstyle Z\kern-0.2em Z$}}}}
\begin{document}
\title{Exact solution of $Z_{2}$ Chern-Simons model on a triangular lattice }
\author{B. Dou\c{c}ot}
\affiliation{Laboratoire de Physique Th\'{e}orique et Hautes \'Energies, CNRS UMR 7589,
Universit\'{e}s Paris 6 et 7, 4, place Jussieu, 75252 Paris Cedex 05 France}
\author{L.B. Ioffe}
\affiliation{Center for Materials Theory, Department of Physics and Astronomy, Rutgers
University 136 Frelinghuysen Rd, Piscataway NJ 08854 USA}

\begin{abstract}
We construct the Hamiltonian description of the Chern-Simons theory with
$Z_{n}$ gauge group on a triangular lattice. We show that the $Z_{2}$ model
can be mapped onto free Majorana fermions and compute the excitation spectrum.
In the bulk the spectrum turns out to be gapless but acquires a gap if a
magnetic term is added to the Hamiltonian. On a lattice edge one gets
additional non-gauge invariant (matter) gapless degrees of freedom whose
number grows linearly with the edge length. Therefore, a small hole in the
lattice plays the role of a charged particle characterized by a non-trivial
projective representation of the gauge group, while a long edge provides a
decoherence mechanism for the fluxes. We discuss briefly the implications for
the implementations of protected qubits.

\end{abstract}
\maketitle

\section{Introduction.}

The challenge of the error free quantum
computation~\cite{Kitaev2002,Childs2002,Preskill1998} resulted in a surge of
interest to many physical systems and mathematical models that were considered
very exotic before. While it is clearly very difficult (if not impossible) to
satisfy the conditions of long decoherence rate and scalability in simple
physical systems\cite{Ioffe2004}, both can be in principle satisfied if
elementary bits are represented by anyons, the particles that indergo
non-trivial transformations when moved adiabatically around each other
(braided)~\cite{Kitaev1997,Mochon2003,Mochon2004}. One of the most famous
examples of such excitations is provided by the fractional Quantum Hall
Effect~\cite{Halperin84,Arovas84}. The difficult part is, of course, to
identify a realistic physical system that has such excitations and allows
their manipulations. This problem should be separated into different layers.
The bottom layer is the physical system itself, the second is the theoretical
model that identifies the low energy processes, the third is the mathematical
model that starts with the most relevant low energy degrees of freedom and
gives the properties of anyons while the fourth deals with construction of the
set of rules on how to move the anyons in order to achieve a set of universal
quantum gates (further lies the layer of algorithms and so on).

One of the most interesting set of problems of the third layer is provided by
the Chern-Simons theories with discrete groups on the lattice: on one hand in
these theories one expects to have a non-local interaction between fluxes that
gives them the anyonic properties, on the other hand they might describe
physics of some solid state arrays. In particular, we have shown very recently
\cite{Doucot2005} that $\nbZ_{2}$ Chern-Simons theory can be realistically
implemented in a Josephson junction array on a square lattice. Furthermore,
the ground state of these physical systems is doubly degenerate but locally
instinguishable so that even a small sized lattice provides a very good
protection of this degeneracy against the external noise. Such protection is
expected to become ideal for the larger lattice sizes if the gap to
excitations is finite. The lack of the exact solution did not allow us to make
definite conclusions on the properties of this model for larger lattice sizes.
However, the analytical solution in limiting cases combined with the extensive
numerics \cite{Dorier2005} indicates that the gap for the low energy
excitations closes in the thermodynamic limit. Practically, this would imply
that protection that can be achieved in such arrays does not grow with the
array size beyond a certain limit. In order to get a better understanding of
the properties of such models we consider here a similar ($\nbZ_{n}$
Chern-Simons) model on a triangular lattice. It turns out that for $n=2$ the
model can be solved exactly by a mapping to Majorana fermions; we find the
gapless spectrum of the photons which acquires a gap if additional
('magnetic') terms are included in the Hamiltonian. We hope that long wave
properties of this model are analogous to the properties of the square lattice
model and thus the problem of the low energy excitations of the latter can be
remedied by the 'magnetic' field term in the Hamiltonian that plays the role
of a tunable chemical potential for fluxes. Because it is difficult to avoid
boundaries in a physical implementations, it is important to understand what
is the effect of edges on the Chern-Simons theory. In particular, from the
view point of the topological protection, it is important to understand
whether they lead to gapless excitations such as edge states in Quantum Hall Effect.

Generally, while the lattice gauge theories without Chern-Simons term are well
studied and understood, much less is known about their Chern-Simons
counterparts. The reason for this is that Chern-Simons term implies coupling
of the charge and flux; on a lattice the charge resides on the sites while the
flux is associated with the plaquettes. Furthermore, one usually wants to
preserve the symmetry of the lattice. To satisfy these criteria one has to
couple charge with the total flux of a few adjacent plaquettes which leads to
novel features absent in the continuous theories. We discuss the details of
this construction in Section II. In particular, for a continuous Abelian gauge
group the Chern-Simons term remains quadratic but becomes non-local in space,
i.e. its Fourier transform acquires momentum dependence; moreover it is zero
for some values of the momenta in the Brillouin zone. This leads to the
appearance of the gapless modes in these theories in the absence of magnetic
energy. Our analysis presented in Section III of the exactly solvable
Chern-Simons theory with $\nbZ_{2}$ group shows that this qualitative picture
holds for the discrete group as well. Finally, in Section IV we study the
boundary effects and find that they are also similar for continuous and
discrete groups. Namely, for continuous groups, the general arguments show
\cite{Witten89} the appearance of a non-gauge invariant matter field on the
boundary in Chern-Simons theories, similarly the exact solution of $\nbZ_{2}$
model shows that the same phenomena happens in discrete models as well.

\section{Model with $\nbZ_{n}$ symmetry}

\begin{figure}[h]
\includegraphics[width=2.0in]{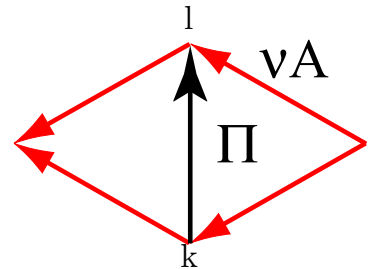}\caption{(Color online) The
orientation convention used in the construction of a Chern-Simons model on a
triangular lattice. The thick (black)\ line represents a $\Pi_{kl}$ operator,
conjugated to the vector potential attached to the link. The arrow is required
since \mbox{$\Pi_{kl}=-\Pi_{lk}$}. The gray (red) lines keep track of signs of
$\nu(kl;lm)$ coefficients which appear in the phase-factors imposed by the
Chern-Simons term in the definition of local electric operators $\mathcal{E}%
_{kl}^{\pm}$ and field translation operators $\mathcal{T}_{kl}^{\pm}$. Gray
(red) lines are oriented from site $m$ to site $n$ whenever the coefficient
$\nu(kl;mn)$ is positive.}%
\label{link}%
\end{figure}

Let us first begin to discuss the construction of the Chern-Simons model with
$\nbZ_{n}$ symmetry on the triangular lattice. The main element of this
construction are the expressions for the electric field operators,
$\mathcal{E}_{kl}^{\pm}$ and the local field translation operators
$\mathcal{T}_{kl}^{\pm}$.\cite{Doucot2005} Both should preserve the symmetry
of the lattice; furthermore, the expression for the electric field operators,
$\mathcal{E}_{kl}^{\pm}$ \ should be gauge invariant while those for
$\mathcal{T}_{kl}^{\pm}$ should preserve the electric fields operators. By
analogy with our previous discussion for the square lattice~\cite{Doucot2005},
we write:
\begin{align*}
\mathcal{E}_{kl}^{\pm}  &  =\exp\left(  \mp i\frac{2\pi}{n}\left(  \Pi
_{kl}-\sum_{(mn)}\nu(kl;mn)A_{mn}\right)  \right) \\
\mathcal{T}_{kl}^{\pm}  &  =\exp\left(  \mp i\frac{2\pi}{n}\left(  \Pi
_{kl}+\sum_{(mn)}\nu(kl;mn)A_{mn}\right)  \right)
\end{align*}
Here $A_{kl}$ and $\Pi_{mn}$ are (discrete) gauge potential and canonical
conjugate momentum on lattice link that satisfy usual canonical commutation
relations:
\[
\lbrack A_{kl},\Pi_{mn}]=i\delta_{(kl),(mn)}%
\]
where $(kl)$ and $(mn)$ denote oriented links on the triangular lattice.
Reversing the orientation of a link changes the sign of the corresponding
variables $A$ and $\Pi$. In the case of a $\nbZ_{n}$ symmetry, the local
vector potentials $A_{kl}$ are constrained to be integer multiples of $2\pi
/n$. The Chern-Simons coefficient $\nu(kl;mn)$ is zero unless $(mn)$ is one of
the four neighbor links of $(kl)$ as illustrated on Fig.~\ref{link}. In this
case, $\nu(kl;mn)=\nu$, if the link $(mn)$ is oriented from right to left for
an observer standing on link (kl) and looking towards site $l$. For the
converse relative orientation, $\nu(kl;mn)=-\nu$

The main consequence of these definitions is that the commutation relations of
electric field operators on nearby links are modified by phase factors. For
$(kl)$ and $(mn)$ oriented as on Fig.~\ref{link}, we have:
\begin{equation}
\mathcal{E}_{kl}^{+}\mathcal{E}_{mn}^{+}=\exp\left(  i(\frac{2\pi}{n})^{2}%
2\nu\right)  \mathcal{E}_{mn}^{+}\mathcal{E}_{kl}^{+}%
\end{equation}
Similarly, the local field translation operators satisfy:
\begin{equation}
\mathcal{T}_{kl}^{+}\mathcal{T}_{mn}^{+}=\exp\left(  -i(\frac{2\pi}{n}%
)^{2}2\nu\right)  \mathcal{T}_{mn}^{+}\mathcal{T}_{kl}^{+}%
\end{equation}
We shall focuss here on the case where these phase factors are equal to $-1$.
This requires $\nu$ to be of the form:
\begin{equation}
\nu=\pi(\frac{n}{2\pi})^{2}(m+\frac{1}{2}) \label{constnu}%
\end{equation}
where $m$ is an integer.
\begin{figure}[h]
\includegraphics[width=2.0in]{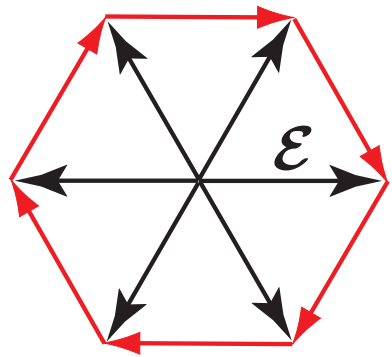}\caption{(Color online)
Construction of the local gauge transformation generator based at the central
site, see Eq.~(\ref{defgauge}). Thick full links correspond to electric field
operators $\mathcal{E}_{kl}^{+}$, and grey (red) oriented links to phase
factors, $\exp(i\frac{8\pi\nu}{n}A_{mn})$, whose total contribution is equal
to $\exp(-i\frac{8\pi\nu}{n}\Phi_{k})$, where $\Phi_{k}$ is the total flux
(counted counterclockwise) through the hexagon.}%
\label{site}%
\end{figure}

To construct the $\nbZ_{n}$ model, we start from the system where local vector
potentials can be arbitrary (and in particular unbounded) integer multiples of
$2\pi/n$, and keep then only the subspace of periodic states $|\Psi\rangle$
which satisfy:
\begin{equation}
(\mathcal{T}_{kl}^{+})^{n}|\Psi\rangle=|\Psi\rangle
\end{equation}
In general, this condition is not compatible with the possibility to perform
arbitrary local gauge transformations, because the $\mathcal{T}_{kl}^{+}$ no
longer commute with generators of local gauge transformations $U_{k}$ defined
as:
\begin{equation}
U_{k}=\prod_{l}^{(k)}\mathcal{T}_{kl}^{+}=\prod_{l}^{(k)}\mathcal{E}_{kl}%
^{+}\exp(-i\frac{2\pi}{n}4\nu\Phi_{k}) \label{defgauge}%
\end{equation}
where $\Phi_{k}=\sum_{hex}A_{mn}$ is the total flux through the loop composed
by the six first neighbors of site $k$ and oriented counterclockwise, as
illustrated on Fig.~\ref{site}. This is an integer multiple of $2\pi/n$. Note
that some care is required in chosing the order of the six operators involved
in the above products. A convenient choice is to lump together each link
$(kl)$ with its opposite $(kl^{\prime})$. This convention is sufficient to
specify the total phase factor because $\mathcal{T}_{kl}^{+}$ operators on all
but adjacent links commute (so that $[\mathcal{T}_{kl}^{+},\mathcal{T}%
_{kl^{\prime}}^{+}]=0$) while different pairs $\mathcal{T}_{kl}^{+}%
\mathcal{T}_{kl^{\prime}}^{+}$ commute because of the cancellation of the
phase factors for any value of $\nu$. So, this choice yields three pairwise
products which commute among themselves. With the notations of Fig.~\ref{site}%
, we have:
\begin{equation}
U_{k}\mathcal{T}_{mn}=\exp\left(  -i(\frac{2\pi}{n})^{2}4\nu\right)
\mathcal{T}_{mn}U_{k}%
\end{equation}
Therefore, the $\nbZ_{n}$ periodicity conditions are compatible with local
gauge invariance provided the quantity $(2\pi/n)^{2}4\nu n$ is an integer
multiple of $2\pi$, which is clearly the case for $\nu$ chosen according to
Eq.~(\ref{constnu}). Another important consequence of this choice is the very
simple connection between generators of gauge transformations and local
electric fields:
\begin{equation}
U_{k}=\prod_{l}^{(k)}\mathcal{E}_{kl}^{+} \label{gaugelec}%
\end{equation}
A last consequence is that for any \emph{even} value of $n$:
\begin{equation}
(\mathcal{E}_{kl}^{+})^{n}=(\mathcal{T}_{kl}^{+})^{n} \label{elecn}%
\end{equation}
Once this set of basic operators has been defined, the natural gauge-invariant
Hamiltonian is constructed as a sum of two terms: the first one involves local
electric field operators on individual links and the second one the local
magnetic fluxes around plaquettes. Specifically:
\begin{equation}
H=-\frac{1}{2}\sum_{(k,l)}(\mathcal{E}_{kl}^{+}+\mathcal{E}_{kl}^{-}%
)+\sum_{(j,k,l)}f(\Phi(j,k,l)) \label{H}%
\end{equation}
The first sum is over links $(k,l)$ and the second one over elementary
triangular plaquettes $(j,k,l)$ which are oriented counterclockwise. At this
stage, $f$ may be any function and the flux
\mbox{$\Phi(j,k,l)=A_{ij}+A_{jk}+A_{kl}$}. Note that we are not dealing here
with the pure Chern-Simons theory, but rather with a discrete analogue of a
Maxwell-Chern Simons theory. Indeed, the former has only a small Hilbert space
which dimension is independent of the system size, and a vanishing
Hamiltonian. In the pure Chern-Simons theory, only fluxless configurations are
allowed, unless the ambiant space has a non-trivial topology. The Hilbert
space of the Maxwell Chern-Simons theory is much larger, and is much more
likely to correspond to the low energy sector of a real physical system. The
ground-state sector of this model is then expected to be described by a pure
Chern-Simons model.

\section{The $n=2$ case}

From now on, we assume $n=2$. We shall map the subspace of gauge invariant and
$2\pi$ periodic states (the \textquotedblleft physical
subspace\textquotedblright) on a system of Majorana fermions attached to the
plaquettes of the original triangular lattice. The periodicity condition
implies, thanks to eq.~(\ref{elecn}) that physical states $|\Psi\rangle$
satisfy:
\begin{equation}
(\mathcal{E}_{kl}^{+})^{2}|\Psi\rangle=|\Psi\rangle
\end{equation}
We may then set:
\begin{equation}
\mathcal{E}_{kl}^{+}=\mathcal{E}_{kl}^{-}\equiv\mathcal{E}_{kl}%
\end{equation}
on this physical subspace.
\begin{figure}[h]
\includegraphics[width=3.0in]{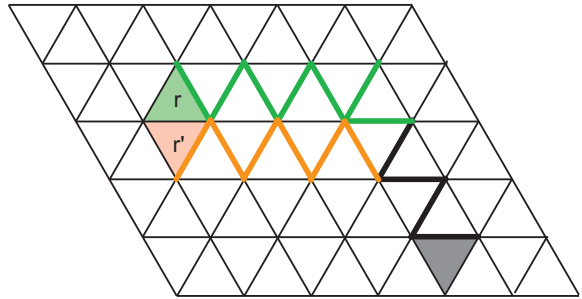}\caption{(Color online) Strings
of electric field operators (thick lines), $S_{\mathbf{r}}$ and
$S_{\mathbf{r^{\prime}}}$ that create or remove $\nbZ_{2}$ fluxes in the
plaquettes indicated by the shaded triangles. For two arbitrary fluxes the
corresponding strings have a common part (full lines) and individual parts
(gray lines). Two individual parts form a contour that connects two fluxes.
Each of these three substrings anticommutes with all others that results in
the anticommutation of the flux creation operators (see text) }%
\label{string}%
\end{figure}
As for the square lattice, it is fruitful to introduce string-like operators
$S_{\mathbf{r}}$ which create or remove a $\nbZ_{2}$ flux on a plaquette
labelled by a site $\mathbf{r}$ on the hexagonal dual lattice. This operator
involves the product of $\mathcal{E}_{kl}$ on a contour that starts from a
specific bond of the lattice, as shown on Fig. \ref{string}. For the gauge
invariant states all such contours are equivalent, so this contour can be
chosen arbitrarily. For instance, one can choose the contours that first go up
and then left toward the plaquette $r$. Using the fact that electric operators
on nearby links anti-commute, we have:
\begin{equation}
S_{\mathbf{r}}^{2}=(-1)^{l(\mathbf{r})-1} \label{S^2}%
\end{equation}
where $l(\mathbf{r})$ is the length of the contour $\mathbf{r}$, namely it is
the total number of electric operators involved in the construction of
$S_{\mathbf{r}}$. Because a gauge transformation changes the length of the
contour by an even number, the right-hand side of Eq.~(\ref{S^2}) depends only
on the parity, $[\mathbf{r}]\equiv(l(r)-1)\operatorname{mod}2$, of the
plaquette $\mathbf{r}$ and not on the contour leading to it. Note that
$S_{\mathbf{r}}$ is not always a hermitian operator, since
\begin{equation}
S_{\mathbf{r}}^{\dagger}=(-1)^{[\mathbf{r}]}S_{\mathbf{r}}%
\end{equation}
The commutation rules obeyed by these string operators are the following:
\begin{equation}
\{S_{\mathbf{r}},S_{\mathbf{r^{\prime}}}\}=2\delta(\mathbf{r}%
,\mathbf{r^{\prime}})(-1)^{[\mathbf{r}]} \label{S_rS_r'}%
\end{equation}
This crucial property which exhibits a clear fermionic behavior can be proved
in two ways. First, one can deform the contours so that they overlap between
point $0$ and $\mathbf{r}$. Then $S_{\mathbf{r^{\prime}}}=S_{\mathbf{r^{\prime
},r}}S_{\mathbf{r}}$ where $S_{\mathbf{r^{\prime},r}}$ is the contour
beginning at $\mathbf{r}$ and ending at $\mathbf{r}^{\prime}\mathrm{.}$ This
contour $S_{\mathbf{r^{\prime},r}}$ contains exactly one electric field
operator namely its first one, which anticommutes with the last electric field
operator entering in $S_{\mathbf{r}}$, so operators $S_{\mathbf{r^{\prime},r}%
}$ and $S_{\mathbf{r}}$ anticommute which proves (\ref{S_rS_r'}).
Alternatively, one can write the original $S$ operators as products
$S_{\mathbf{r}}=\widetilde{S}_{\mathbf{r}}S_{0}$ and $S_{\mathbf{r}^{\prime}%
}=\widetilde{S}_{\mathbf{r}^{\prime}}S_{0}$ \ where $S_{0}$ is the product of
electric field operators on the common part of the contour leading to
plaquettes $\mathbf{r}$ and $\mathbf{r}^{\prime}$, see Fig.~\ref{string}. The
operators $\widetilde{S}_{\mathbf{r}}$ and $S_{0}$ have exactly one pair of
nearest neighboring electric fields, so they anticommute: $\{\widetilde
{S}_{\mathbf{r}^{\prime}},S_{0}\}=\{\widetilde{S}_{\mathbf{r}},S_{0}\}=0$.
Thus, $S_{\mathbf{r}}S_{\mathbf{r^{\prime}}}=-S_{0}^{2}\widetilde
{S}_{\mathbf{r}}\widetilde{S}_{\mathbf{r}^{\prime}}$ and $S_{\mathbf{r}%
^{\prime}}S_{\mathbf{r}}=-S_{0}^{2}\widetilde{S}_{\mathbf{r}^{\prime}%
}\widetilde{S}_{\mathbf{r}}$, further, noticing that the operators
$\widetilde{S}_{\mathbf{r}}\widetilde{S}_{\mathbf{r}^{\prime}}$ contain
exactly one pair of anticommuting electric fields, we get (\ref{S_rS_r'}).

A very important property of this model restricted to its physical subspace
defined above is that local electric operators on link $(kl)$ are simply
related to bilinear expressions of the form $S_{\mathbf{r}}%
S_{\mathbf{r^{\prime}}}$ where the plaquettes $\mathbf{r}$ and
$\mathbf{r^{\prime}}$ are located on both sides of the link $(kl)$. If
$\mathbf{r}$ and $\mathbf{r^{\prime}}$ are nearest neighbor plaquettes, we
have:
\begin{equation}
S_{\mathbf{r}}=\mathcal{E}_{kl}S_{\mathbf{r^{\prime}}} \label{horizontalhop}%
\end{equation}
which is a direct consequence of the definition of string operators. Thus,
\begin{equation}
\mathcal{E}_{kl}=(-1)^{[\mathbf{r}^{\prime}]}S_{\mathbf{r}}%
S_{\mathbf{r^{\prime}}} \label{E_kl_1}%
\end{equation}

The last stage is to introduce Majorana fermions $\chi_{\mathbf{r}}$ related
to string operators by:
\begin{equation}
S_{\mathbf{r}}=i^{[\mathbf{r}]}\chi_{\mathbf{r}}%
\end{equation}
With this definition, it is easy to check that:
\begin{align}
\chi_{\mathbf{r}}^{\dagger}  &  =\chi_{\mathbf{r}}\\
\{\chi_{\mathbf{r}},\chi_{\mathbf{r^{\prime}}}\}  &  = 2 \delta_{\mathbf{r}%
,\mathbf{r^{\prime}}}%
\end{align}
Eq.~(\ref{E_kl_1}) becomes:
\begin{equation}
\mathcal{E}_{kl}=i\chi_{\mathbf{r}}\chi_{\mathbf{r^{\prime}}}
\label{E_kl_Majorana}%
\end{equation}
if site $\mathbf{r}^{\prime}$ is even ($[\mathbf{r}^{\prime}]=0$) and site
$\mathbf{r}$ is therefore odd.

The electrical part of the Hamiltonian thus maps into the hopping of Majorana
fermions. The full Hamiltonian includes also the magnetic field part. In case
of $\nbZ_{2}$ model it is given by the operator that takes two values
depending on the flux in a given plaquette, so a general function $f(\Phi)$
from (\ref{H}) is reduced to a linear function $f(\Phi)=\mu\Phi$. Writing the
Majorana fermion as a sum of two usual fermion operators
\mbox{$\chi=c+c^{\dagger}$}, we see that the total fermion number $c^{\dagger
}c$ changes by $\pm1$ by the flux creation operator, so in the fermion
language the full Hamiltonian becomes:
\[
H=-i\sum_{(\mathbf{r},\mathbf{s})}(c_{\mathbf{r}}+c_{\mathbf{r}}^{\dagger
})(c_{\mathbf{s}}+c_{\mathbf{s}}^{\dagger})+\mu\sum_{\mathbf{r}}c_{\mathbf{r}%
}^{\dagger}c_{\mathbf{r}}%
\]
where the indices $\mathbf{r},\mathbf{s}$ run over the sites of the dual
(hexagon) lattice and the first sum goes over all nearest neighbours on this
lattice. The spectrum of the excitations
\begin{equation}
E(k)=\sqrt{\mu^{2}+|t(k)|^{2}}\pm|t(k)| \label{E(k)}%
\end{equation}
where $t(k)$ is spectrum of fermions on the honeycomb lattice with a purely
nearest neighbour hopping Hamiltonian $H_{t}=\sum_{(\mathbf{r},\mathbf{s}%
)}c_{\mathbf{r}}^{\dagger}c_{\mathbf{s}}$. To compute $t(k)$ we choose
elementary cell consisting of two sites, for instance the ones that belong to
the same vertical bond. In momentum space the Hamiltonian becomes a matrix
$H=c_{a,k}^{\dagger}K_{ab}c_{b,k}$ with
\[
K=\left(
\begin{array}
[c]{cc}%
0 & 1+e^{i\mathbf{k\xi}}+e^{i\mathbf{k\eta}}\\
1+e^{-i\mathbf{k\xi}}+e^{-i\mathbf{k\eta}} & 0
\end{array}
\right)
\]
where $\mathbf{\xi=(}\frac{\sqrt{3}}{2},\frac{3}{2})$, $\mathbf{\eta=(-}%
\frac{\sqrt{3}}{2},\frac{3}{2})$ are unit vectors of the lattice of vertical
bonds. We get
\[
t(k)=\sqrt{1+4\cos\left(  \frac{\sqrt{3}}{2}k_{x}\right)  \cos\left(  \frac
{3}{2}k_{y}\right)  +4\cos^{2}\left(  \frac{\sqrt{3}}{2}k_{x}\right)  }%
\]

In the absence of magnetic term the spectrum has dispersionless mode and a
dispersive one which has zero only at the isolated Fermi points $(\pm
4\pi/(3\sqrt{3}),0)$, $(\pm2\pi/(3\sqrt{3}),2\pi/3)$. Near the Fermi point the
spectrum is linear $E(k)=3|k|$. In the presence of magnetic term two things
happen: the dispersive mode acquires a gap $2\mu$, near the Fermi point the
spectrum becomes massive $E(k)=\sqrt{(3/2)^{2}|k|^{2}+\mu}+(3/2)|k|$ and the
dispersionless mode acquires $k$-dependence with the gap $\sqrt{\mu^{2}+9}-3.$

The presence of a dispersionless band for $\mu=0$ is directly connected to the
presence of a set of local symmetries in the model. Indeed, if we introduce
the Majorana operators \mbox{$\widetilde{\chi}=i(c-c^{\dagger})$}, they
satisfy:
\begin{align}
\widetilde{\chi}_{\mathbf{r}}^{\dagger}  &  =\widetilde{\chi}_{\mathbf{r}}\\
\{\widetilde{\chi}_{\mathbf{r}},\widetilde{\chi}_{\mathbf{r^{\prime}}}\}  &  =
2 \delta_{\mathbf{r},\mathbf{r^{\prime}}}\\
\{\chi_{\mathbf{r}},\widetilde{\chi}_{\mathbf{r}}\}  &  = 0
\end{align}
As a result, $\widetilde{\chi}_{\mathbf{r}}$ commutes with $H$ for any
$\mathbf{r}$, therefore providing a set of non-commuting local symmetries
which are destroyed by the presence of a magnetic term $\mu$.

\section{Edge States}

\begin{figure}[h]
\includegraphics[width=3.0in]{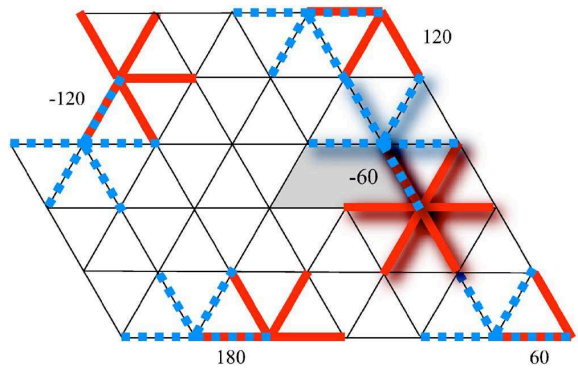}\caption{(Color online)
Commutation relations between gauge generators attached to boundary sites in
the $\nbZ_{2}$ Chern-Simons model. We show various possible edge shapes and
indicate the corresponding opening angle expressed in degrees. For each
situation, we display the gauge generators for a pair of nearest-neighbor
sites. These generators involve the electric operators originating from a
given site, and are depicted either as full (red) thick lines or dashed (blue)
thick lines. For all situations appearing on the outer boundary of the
lattice, these gauge generators are found to \emph{anticommute}, which induces
degenerate multiplets for the gauge symmetry. An exception to this rule is the
$-60$ cusp shown on an inner hole for which the two nearby generators
commute.}%
\label{EdgeStates}%
\end{figure}

An interesting feature of the present model is that it exhibits edge states
localized around boundaries. This is known to be a general property of
Chern-Simons theories~\cite{Witten89}, but an advantage of a lattice gauge
theory with a discrete gauge group over a model defined on continuous space is
to yield a \emph{finite dimensional} Hilbert space. Therefore, we do not
require any sophisticated analysis to specify boundary conditions. In the
presence of a boundary, the definition of basic operators $\mathcal{E}%
_{kl}^{\pm}$ and $\mathcal{T}_{kl}^{\pm}$ for a link $kl$ along the boundary
is the same as before, with the exception that the associated phase-factor
only involves a smaller set of vector potentials $A_{mn}$ attached to the
neighboring links lying \emph{inside} the system. It is then easy to check
that this does \emph{not} modify the basic commutation relations between these
operators. In particular, we shall still concentrate here on the $n=2$ case
where nearby local field translation operators anticommute. The gauge
generators $U_{k}$ located at site $k$ is still the product of all
$\mathcal{T}_{kl}^{\pm}$ operators such that $kl$ lies inside the system. As
illustrated on Fig.~\ref{EdgeStates} this implies that $U_{k}$ and $U_{l}$
still commute when at least site $k$ or $l$ is not on the boundary, but they
now \emph{anticommute} if $k$ and $l$ are nearest neighbor sites both located
on the boundary. This holds for most possible shapes of the boundary, with few
exceptions which are depicted on Fig.~\ref{EdgeStates}. The striking
consequence of this is that it is no longer possible to diagonalize
simultanenously all local gauge generators $U_{k}$ belonging to the same
boundary. In particular, gauge singlets no longer exist since they are
replaced by degenerate multiplets corresponding to \emph{projective}
representations of the gauge group associated to boundary sites.

Let us now describe these multiplets in some detail. For this, we consider a
finite triangular lattice with an outer boundary, and possibly with some inner
holes, each of them bringing its own boundary. We assume that these holes are
not too close from each other, so that local gauge generators associated to
sites belonging to two different boundaries always commute. This assumption
allows us to treat each boundary separately from the others. To simplify the
discussion, we shall consider first the edge without $-60$ degrees turns that
contains an even number $L$ of sites, which will be labelled by indices $n$
running from 1 to $L$. We suppose that fluxes in each plaquette and through
each inner hole have been fixed, thereby concentrating on the degrees of
freedom attached to gauge transformations. We may then associate to our
boundary a Hilbert space containing $2^{L}$ independent states. In order to
construct irreducible representations of the boundary gauge-group, we have to
find a maximal subset of mutually commuting generators, which are then
diagonalized simultaneously. Such set may be chosen as follows: take first
generators $U_{2n}$ on even sites, and add then a global gauge generator
$\prod_{n=1}^{L}U_{n}$ for the boundary. This yields $2^{1+L/2}$ possible sets
of quantum numbers corresponding to:
\begin{align}
U_{2n}|\Psi\rangle &  =\tau_{2n}|\Psi\rangle\\
\prod_{n=1}^{L}U_{n}|\Psi\rangle &  =\tau|\Psi\rangle
\end{align}
where eigenvalues $\tau_{2n}$ and $\tau$ can be $\pm1$. We may still interpret
$\tau$ as the total $\nbZ_{2}$ charge of matter induced on the boundary by the
Chern-Simons term. Starting from a common eigenstate $|\Psi\rangle$ as above,
applying $U_{2n+1}$ produces a new eigenstate in which $\tau_{2n}$ and
$\tau_{2n+2}$ have changed sign simultaneously. Note that we have:
\begin{equation}
\prod_{n=1}^{\frac{L}{2}}U_{2n-1}|\Psi\rangle=\tau\prod_{n=1}^{\frac{L}{2}%
}\tau_{2n}|\Psi\rangle
\end{equation}
so only $(L/2)-1$ generators located on odd sites act independently. We
therefore generate a $2^{(L/2)-1}$ dimensional irreducible multiplet, so the
Hilbert space attached to the boundary splits into $2^{(L/2)+1}$ such
degenerate subspaces.

The above construction provides a basis in each multiplet for which $U_{2n}$
is represented by a diagonal matrix, whereas $U_{2n+1}$'s play the role of
raising or lowering operators. A convenient way to visualize these multiplets
is to describe them in terms of an effective spin 1/2 model, attached to even
boundary sites. This correspondence is given by:
\begin{align}
U_{2n}  &  =\tau_{2n}^{z},\;\;\;(1\leq n\leq\frac{L}{2})\\
U_{1}  &  =\tau_{L}^{x}\tau_{2}^{x}\\
U_{2n-1}  &  =\tau_{2n-2}^{x}\tau_{2n}^{x},\;\;\;(2\leq n\leq\frac{L}{2}-1)\\
U_{L-1}  &  =\tau\left(  \prod_{n=1}^{\frac{L}{2}}\tau_{2n}^{z}\right)
\tau_{L-2}^{x}\tau_{L}^{x}%
\end{align}
A given multiplet corresponds to a fixed eigenvalue for $\prod_{n=1}^{\frac
{L}{2}}\tau_{2n}^{z}$, so that the number of independent Ising spins is only
$(L/2)-1$ as discussed above. Note that these conclusions apply to
sufficiently large holes in the lattice. For very small holes consisting of
two triangles glued together to form a rhombus (and of course for a single
triangle itself) all boundary operators commute. The first pair of
anticommuting operators appears in a trapezoidal hole shown in Fig.
\ref{EdgeStates}. This hole carries an effective spin $1/2$ degree of freedom
because it is possible to diagonalize all gauge generators except one. Note
that in this special case the product $\prod_{n=1}^{5}U_{n}$
\emph{anticommutes} with the two local generators on upper sites, so the
corresponding charge is no longer a conserved quantum number. In larger holes,
such as hexagon, all nearest neighbor gauge generators anticommute. For
instance, the boundary matter of elementary hexagon can be described as an
effective system composed of \emph{two} spins 1/2.

What is the interaction between these new matter-like degrees of freedom and
the fluxons described in the previous section? Suppose first that these
fluxons are not allowed to jump accross boundaries. This corresponds to a
Hamiltonian where the electrical operators attached to boundary links are
removed. The only interaction between the two sets of degrees of freedom
occurs via the usual Aharonov-Casher phase-factor for fluxons going around
inner holes. Each hole generates an effective $\nbZ_{2}$ orbital magnetic
field which produces an adiabatic phase $\phi$ related to the $\nbZ_{2}$
charge by $e^{i\phi}=\tau$. This effect does not disturb the internal states
of the boundaries which remain constants of motion. A stronger interaction
occurs when fluxons are allowed to cross boundaries (that is to disappear from
a boundary placquette). A simple description of this interaction is possible
if a fluxon moves around a closed loop which crosses twice a given boundary.
In this process, the final internal state of the boundary is connected to the
initial one by applying the product of all local boundary gauge generators
which are also enclosed in the closed loop. In the example of a hole with the
shape of an elementary hexagon, these scattering processes can be simply
expressed as products of operators taken in the set
\mbox{$\tau^{z}_{2},\tau^{x}_{2},\tau^{z}_{4},\tau^{x}_{4}$}. In connection
with the idea of topological quantum computation, we may wonder if these well
defined operations can be used as a physical basis for qubits. The problem
with this idea is that local gauge symmetry operators induce transitions
between the various states of this effective spin system attached to the
boundary. Since in any real implementation, noise acts through basically any
local operator, such states cannot be protected from environment-induced
decoherence. If we now consider the effect of either a large boundary, or many
small inner holes, on the fluxon dynamics, we see that these processes tend to
entangle fluxon states with an effective environment associated to boundary
matter, resulting in the absence of long-ranged phase coherence for the
Majorana fermions describing the fluxons. It is also very unlikely that the
model preserves its integrabiblity, as soon as several fluxons and boundary
matter are simultaneously present.

\section{Conclusions}

In this paper, we have solved exactly a $\nbZ_{2}$ gauge theory with a
Chern-Simons term on a triangular lattice by mapping the gauge-invariant
subspace into a free Majorana fermion system. These fermions keep track of the
dynamics of local fluxes. In the absence of an energy cost to create a
$\nbZ_{2}$ flux, the spectrum is very degenerate:\ it exhibits a flat band at
zero energy and a dispersive one with a vanishing energy at isolated points on
the Brillouin zone boundary. This picture is very similar to the solution
found by Kitaev for a quantum spin 1/2 model on the hexagonal
lattice~\cite{Kitaev03}. In this model, he considered nearest-neighbor
interactions of the form $(\mbox{\boldmath $\sigma$}_{i}%
.\mbox{\boldmath $n$}_{ij})(\mbox{\boldmath $\sigma$}_{j}%
.\mbox{\boldmath $n$}_{ij})$ where $\mathbf{n}_{ij}=\mathbf{e}_{x}%
,\mathbf{e}_{y},\mathbf{e}_{z}$ depending on the unit vector joining sites $i$
and $j$. These operators have the same anticommutation properties as the local
fluxon moving operators
\mbox{$\mathcal{E}_{kl}=i\chi_{\mathbf{r}}\chi_{\mathbf{r^{\prime}}}\label{E_kl}$}
introduced in the Chern-Simons lattice gauge model. However, we have not found
a way to establish a one to one mapping between these two models. We do not
see any operator in Kitaev's spin model which could be interpreted as a local
fluxon number, since it would have to anticommute with the three components
$\sigma^{x}$, $\sigma^{y}$, $\sigma^{z}$ of the local spin at any site. In our
gauge theory, fluxons have to be created by non-local operators $S_{\mathbf{r}%
}$ associated to contours, so it does not seem easy to express them in terms
of local spin operators.

The presence of a gapless spectrum in the absence of a magnetic energy term is
reminiscent of the gapless phase found on a square
lattice~\cite{Doucot2005,Dorier2005}. On a square lattice, this $\nbZ_{2}$
Chern-Simons model can be mapped into a spin 1/2 model with anisotropic
exchange interactions of the form $\sigma^{x}_{i}\sigma^{x}_{j}$ or
$\sigma^{z}_{i}\sigma^{z}_{j}$ depending on the orientation of the link $ij$.
Adding a magnetic term in the Chern-Simons theory is then equivalent to
imposing a magnetic field along the $\mathbf{y}$ direction in the spin model.
It would be interesting to check that it also induces a spectral gap.
Unfortunately, such a magnetic field term does not commute with the non-local
symmetry operators associated to rows and columns of the square lattice which
were responsible for the two-fold degeneracy of all energy eigenstates.

Finally, we have shown explicitely that charged matter degrees of freedom are
always induced along edges associated to the external boundary or to inner
holes. These degrees of freedom appear naturally since the presence of an edge
forces the representation of the local gauge symmetry to become projective,
i.e. gauge generators attached to nearest neighbor sites both located along an
edge do not commute. We have shown that these new degrees of freedom provide
static Aharonov-Bohm fluxes for the orbital motion of fluxons, provided the
latter are not allowed to cross edges. Transitions within these degenerate
matter multiplets are induced by processes where a fluxon goes back and forth
accross a boundary. Unfortunately, these multiplets are very sensitive to
external noise acting through local operators, so it is unlikely they may
serve the purpose of designing protected qubits.

It is an immediate extension of this work to add local static charges in the
system and see how they interact with this fluid of Majorana fermions
associated to fluxons. Very likely, the interaction will be of the
Ahoronv-Casher type, which is too weak to induce a confining force for a pair
of opposite charges. So again, we have a direct evidence that a Chern-Simons
term distroys the confining regime expected in gauge theories with small
magnetic energy in the absence of a Chern-Simons
term~\cite{Pisarski86,Affleck89}.

The most important question connected to protected quantum computation is the
construction of similar models for a finite non-Abelian
group~\cite{Mochon2003,Mochon2004}. This is the subject of a forthcoming work.

\textbf{Acknowledgments}

We are thankful to M.V. Feigelman and A. Silva for the critical reading of the
paper. LI is thankful to LPTHE, Jussieu for their hospitality while BD has
enjoyed the hospitality of the Physics Department at Rutgers University. We
thank A. Kitaev for kindly sending us his unpublished work on the honeycomb
lattice which encouraged us to study the Chern-Simons theory on the triangular
lattice. This work was made possible by support from NATO CLG grant 979979,
NSF DMR 0210575.

\end{document}